\documentclass[12pt]{article}
\usepackage{graphicx,amssymb,bbold,bm}
\usepackage{amsmath,amsfonts,epsfig}
\usepackage{chessfss}
\usepackage{braket}
\usepackage[usenames]{color} 
\usepackage{wrapfig}
\usepackage{amsfonts,bm}
\usepackage{epsfig,amsfonts,bm}
\usepackage{tikz}

\newcommand{\p}{\partial}
\newcommand{\pslash}{p\kern-1ex /}
\newcommand{\qslash}{q\kern-1ex /}
\newcommand{\lslash}{l\kern-1ex /}
\newcommand{\sslash}{s\kern-1ex /}
\newcommand{\kaslash}{k_a\kern-2ex /}
\newcommand{\kbslash}{k_b\kern-2ex /}
\newcommand{\Dslash}{{\cal D}\kern-1.5ex /}

\newcommand{\bc}{\overline{c}}

\newcommand{\beqa}{\begin{eqnarray}}
\newcommand{\eeqa}{\end{eqnarray}}

\newcommand{\bpm}{\begin{pmatrix}}
\newcommand{\epm}{\end{pmatrix}}
\newcommand{\bbm}{\begin{bmatrix}}
\newcommand{\ebm}{\end{bmatrix}}

\def\p{\partial}

\def\mS{{\mathbb S}}

\makeatletter

\@addtoreset{equation}{section}
\makeatother
\usepackage{subfig}
\begin{document}


\voffset -0.7 true cm
\hoffset 1.5 true cm
\topmargin 0.0in
\evensidemargin 0.0in
\oddsidemargin 0.0in
\textheight 8.6in
\textwidth 5.4in
\parskip 9 pt
 
\def\Tr{\hbox{Tr}}
\newcommand{\be}{\begin{equation}}
\newcommand{\ee}{\end{equation}}
\newcommand{\bea}{\begin{eqnarray}}
\newcommand{\eea}{\end{eqnarray}}
\newcommand{\beas}{\begin{eqnarray*}}
\newcommand{\eeas}{\end{eqnarray*}}
\newcommand{\nn}{\nonumber}
\font\cmsss=cmss8
\def\C{{\hbox{\cmsss C}}}
\font\cmss=cmss10
\def\bigC{{\hbox{\cmss C}}}
\def\scriptlap{{\kern1pt\vbox{\hrule height 0.8pt\hbox{\vrule width 0.8pt
  \hskip2pt\vbox{\vskip 4pt}\hskip 2pt\vrule width 0.4pt}\hrule height 0.4pt}
  \kern1pt}}
\def\ba{{\bar{a}}}
\def\bb{{\bar{b}}}
\def\bc{{\bar{c}}}
\def\bphi{{\Phi}}
\def\Bigggl{\mathopen\Biggg}
\def\Bigggr{\mathclose\Biggg}
\def\Biggg#1{{\hbox{$\left#1\vbox to 25pt{}\right.\n@space$}}}
\def\n@space{\nulldelimiterspace=0pt \m@th}
\def\m@th{\mathsurround = 0pt}

\begin{titlepage}
\begin{flushright}
{\small OU-HET-1069}
 \\
\end{flushright}

\begin{center}

\vspace{5mm}
{\large \bf {The averaged null energy conditions in even}} \\[1pt] 
\vspace{1mm}
{\large \bf  {dimensional curved spacetimes from AdS/CFT duality}} \\[2pt]

\vspace{6mm}

\renewcommand\thefootnote{\mbox{$\fnsymbol{footnote}$}}
Norihiro Iizuka${}^{\textsymking}$, 
Akihiro Ishibashi${}^{\textsymbishop}$ and 
Kengo Maeda${}^{\textsymqueen}$

${}^{\textsymking}${\small \sl Department of Physics, Osaka University} \\ 
{\small \sl Toyonaka, Osaka 560-0043, JAPAN}

${}^{\textsymbishop}${\small \sl Department of Physics and Research Institute for Science and Technology,} \\   
{\small \sl Kindai University, Higashi-Osaka 577-8502, JAPAN}
 
${}^{\textsymqueen}${\small \sl Faculty of Engineering, Shibaura Institute of Technology,} \\   
{\small \sl Saitama 330-8570, JAPAN} 

{\small \tt 
{iizuka at phys.sci.osaka-u.ac.jp}, {akihiro at phys.kindai.ac.jp},   \\
{maeda302 at sic.shibaura-it.ac.jp}
}

\end{center}


\noindent

We consider averaged null energy conditions (ANEC) for strongly coupled quantum field theories in even (two and four) dimensional curved 
spacetimes by applying the no-bulk-shortcut principle in the context of the AdS/CFT duality. 
In the same context but in odd-dimensions, the present authors previously derived a conformally 
invariant averaged null energy condition (CANEC), which is a version of the ANEC with a certain weight function 
for conformal invariance. In even-dimensions, however, one has to deal with gravitational conformal anomalies, 
which make relevant formulas much more complicated than the odd-dimensional case.  
In two-dimensions, we derive the ANEC by applying the no-bulk-shortcut principle. 
In four-dimensions, we derive an inequality which essentially provides the lower-bound 
for the ANEC with a weight function. For this purpose, and also to get some geometric insights into gravitational conformal anomalies, 
we express the stress-energy formulas in terms of geometric quantities such as the expansions of boundary null geodesics 
and a quasi-local mass of the boundary geometry.  
We argue when the lowest bound is achieved and also discuss when the averaged value of the null energy can be negative, 
considering a simple example of a spatially compact universe with wormhole throat.

\end{titlepage}

\setcounter{footnote}{0}
\renewcommand\thefootnote{\mbox{\arabic{footnote}}}

\tableofcontents


\section{Introduction} 

The null energy condition~(NEC) is the key to prove a number of theorems in general relativity, such as 
the singularity theorems, topological censorship, and positive energy theorems. Although it is satisfied for typical classical matter fields,  as a locally formulated condition, the NEC can be violated by quantum effects, such as Casimir effects in 
spacetime with a compact spatial section.

In many of its applications, the NEC can be replaced by the averaged null energy condition~(ANEC), 
\begin{align}
\label{simpleANEC}
\int^{+\infty}_{-\infty}  \braket{T_{\mu\nu}} \,l^\mu l^\nu d\lambda\ge 0 ,
\end{align}
for an achronal null geodesic with tangent vector $l^\mu$, where $\lambda$ is the affine parameter along the null geodesic
and $\braket{T_{\mu\nu}}$ is the stress-energy tensor.   
This states that the integral of the null energy over a complete null geodesic 
cannot be negative, as first realized in \cite{Tipler1978}, proved for some cases \cite{WaldYurtsever91}, 
and improved, e.g., by \cite{GrahamOlum07, FlanaganWald96}.  
In Minkowski spacetime, the (achronal) ANEC has recently been proven for general quantum fields~\cite{Faulkner:2016mzt, Hartman:2016lgu}. The ANEC in the Minkowski background was also shown for strongly coupled conformal fields~\cite{Kelly:2014mra} in the context of the AdS/CFT duality~\cite{Maldacena:1997re}. 

In curved spacetimes, however, the ANEC has not been fully tested yet. 
Although it has recently been shown to hold in the maximally 
symmetric curved (i.e., de Sitter and Anti-de Sitter) spacetimes~\cite{Rosso2019} (see also~\cite{Rosso2020} 
for a highly symmetric but non-trivial case), 
the ANEC can in fact be violated for more general cases such as 
a conformally coupled scalar field in $4$-dimensional curved, conformally flat spacetime~\cite{Urban:2009yt}. 
An example of  the ANEC violation in curved spacetimes was also shown in strongly coupled field theory 
in the context of the AdS/CFT duality~\cite{IMM:2019}.

A violation of energy conditions is closely related to causal pathology such as 
the occurrence of naked singularities and/or causality violations. 
This, in turn, suggests that a sensible causality requires a certain energy condition to be satisfied.  
In the AdS/CFT context, a firmer basis of physically reasonable causal interactions between the bulk and boundary 
field theory is provided by imposing the ``no-bulk-shortcut condition," which asserts that no bulk causal curve 
can travel faster than the boundary achronal null geodesics. 
This assertion was precisely formulated and shown by Gao and Wald~\cite{Gao:2000ga}, 
assuming that there are no pathological behavior such as naked singularity formation in the bulk and the boundary. Conversely, 
if the no-bulk-shortcut condition is violated, a naked singularity must appear in the bulk~\cite{IMM:2019}. 
Thus, the no-bulk-shortcut condition is essential to characterize the bulk-boundary causality relation in the AdS/CFT duality.  
In fact, the holographic proof of the ANEC in Minkowiski spacetime \cite{Kelly:2014mra} exploits this condition. 

In the previous paper~\cite{IMM:2019}, the present authors applied 
the no-bulk-shortcut condition in the context of AdS/CFT duality
where the boundaries are $d=3$ and $d=5$ static spatially compact universes, and derived 
the conformally invariant averaged null energy condition (CANEC), 
\begin{align}
\label{CANEC_35}
\int^{\lambda_+}_{\lambda_-} \left(\eta(\lambda) \right)^d\, \braket{T_{\mu\nu}} \,l^\mu l^\nu d\lambda\ge 0 .
\end{align}
Here, $\eta$ is the $d$-dimensional ($d = 3, 5$) boundary Jacobi field of the boundary null geodesic congruence, 
representing the separation of points between the two adjacent null geodesics on the boundary.  
$\lambda_{\pm}$ are conjugate points (focal points) of it.  The formula~\eqref{CANEC_35} is consistent with 
the Minkowski ANEC since in flat spacetime, the Jacobi field becomes constant and focal points are 
$\lambda_{\pm} = \pm \infty$. Similarly in the case of maximally symmetric boundary spacetime, 
Eq.~\eqref{CANEC_35} reduces to ANEC in Eq.~\eqref{simpleANEC} and  
this agrees with the condition derived on the Einstein-static cylinder from field theoretic point of view~\cite{Rosso2019}.

In even-dimensions, however, the boundary conformal field theories in general involve conformal anomalies, 
which make relevant formulas much more complicated than those in odd-dimensions, and 
it is far from obvious if one can generalize in any reasonable way the notion of the CANEC to the even-dimensional case. 
In this paper, we apply the holographic method of our previous paper~\cite{IIM:2020} to the case of even-dimensional boundary spacetimes. 
The main result in previous paper~\cite{IMM:2019} is that ANEC must involve the appropriate weight function by the 
Jacobi field.  
Therefore, again assuming the existence of the holographic bulk duals and also the no-bulk-shortcut principle, 
we derive ANEC in two-dimension and obtain an inequality for ANEC with an appropriate weight function in four dimension. 
This results in providing the lower-bound for the ANEC with a weight function in four-dimensional curved spacetimes. These are our main results in this paper.  However, we will postpone to proving the conformal invariance of our formula due to the complication of conformal anomaly.

The starting point of our holographic method is the Fefferman-Graham~(FG) expansion of  
$(d+1)$-dimensionial asymptotically AdS bulk metric, 
\begin{align}
\label{FG_coordinate}
& g_{ab} dx^a dx^b=\frac{1}{z^2} \left( dz^2+g_{\mu\nu}(x,z)dx^\mu dx^\nu \right) := \frac{1}{z^2} \hat{g}_{ab}\nonumber \\
& g_{\mu\nu}(z,x)=g_{(0)\mu\nu}(x)+z^2g_{(2)\mu\nu}(x)+\cdots + z^d g_{(d)\mu\nu}(x)+h_{(d)\mu\nu}\,z^{d}\ln z^2+\cdots,  
\end{align}
where $d\ge 2$ and $h_{(2)\mu\nu}=0$. $ \hat{g}_{ab}$ is the rescaled 
bulk spacetime metric which we  will use later.  According to the formula \cite{deHaro:2000vlm}, 
the stress-energy tensor $\braket{T_{\mu\nu}}$ in $d$-dimensional boundary field theory 
is given by these expansion coefficients. When $d$ is odd, $\braket{T_{\mu \nu}}$ is simply proportional to 
$g_{(d) \mu \nu}$, while when $d$ is even, there appears an additional term $X_{\mu \nu}$, 
which corresponds to the conformal anomalies of the boundary CFT and makes the formulas significantly involved. 
In the $d=2$ case, $X_{\mu \nu}$ is in proportion to $g_{(0)\mu \nu}$ and therefore 
the null energy $\braket{T_{\mu\nu}}l^\mu l^\nu$ with $l^\mu$ being any null vector is simply given by 
$g_{(2)\mu\nu}l^\mu l^\nu$. 
This fact helps us to control the behavior of relevant bulk and boundary null geodesics in terms only of 
the boundary null energy $\braket{T_{\mu\nu}}l^\mu l^\nu$ and enables us to derive the ANEC in general curved spacetime. 
This includes the ANEC on the complete null geodesic generators on both the $1+1$-dimensional cosmological and black hole horizons. 

In the $d=4$ case, the stress-energy tensor is composed of the coefficient $g_{(4)\mu\nu}$ and the 
addtional term $X_{\mu \nu}$ nonlinear to the curvature 
tensor, reflecting the conformal anomalies~\cite{deHaro:2000vlm}. 
In this case, we derive an energy inequality of the form in which a weighted average of the null energy 
$\braket{T_{\mu\nu}}l^\mu l^\nu$ is bounded from below by boundary geometric quantities, such as 
the expansions of boundary null geodesics, and the quasi-local mass of the boundary spacetime. 
We also show that the equality holds for the defomed global vacuum AdS spacetime with linear perturbations. 
This suggests that the minimum of the averaged null energy $\braket{T_{\mu\nu}}l^\mu l^\nu$ is determined by the boundary 
physical quantities such as the expansions of null geodesics and a quasi-local mass, besides the boundary Ricci tensor. 
We find that the minimum can be {\it negative} for some type of spatially compact 
universe~(see also Refs.~\cite{Hickling:2015tza, Fischetti:2016vfq}).  
 
In the next section, we briefly recall the no-bulk-shortcut condition and holographic stress-energy formulas. 
Then in Sec.~\ref{sec:d=2} we derive the ANEC in the $d=2$ both spatially compact and non-compact universes. In Sec.~\ref{sec:d=4},  
we derive the inequality that the averaged value of the null energy 
$\braket{T_{\mu\nu}}l^\mu l^\nu$ with an appropriate weight is bounded 
from below in $d=4$ dimensional spatially compact universe. Then in Sec.~\ref{sec:mim:energy}, we examine when the equality holds 
in general defomed global AdS spacetime. 
In Sec.~\ref{sec:ex:negative}, we supply a curved boundary example in which the ANEC is violated.  
Sec.~\ref{sec:summary} devotes to summerize our results.

\section{No-bulk-shortcut and boundary stress-energy}
\label{sec:noshortcut}
We would first like to recall the statement of the no-bulk-shortcut principle of Gao-Wald~\cite{Gao:2000ga} 
and some basic formulas for holographic renormalized stress-energy tensor~\cite{deHaro:2000vlm}.

We are concerned with, as our bulk spacetime, a $d+1$-dimensional asymptotically locally anti-de Sitter vacuum spacetime $(M_{d+1},g_{ab})$ 
with conformal boundary $\partial M$.  
Consider any pair of boundary two points, $p, q \in \partial M$, which are connected by an archronal null geodesic $\gamma$ 
lying in $\partial M$~(without loss of generality, we assume that $q$ is located to the future of $p$ in $\partial M$). 
Suppose there exists a timelike curve in the bulk $M_{d+1}$ which anchors to these boundary two points $p$ and $q$. 
Then, there must be another bulk causal curve which connects $p \in \partial M$ and a boundary point $r \in \partial M$ 
which is strictly past to $q \in \partial M$. 
In such a case, $M_{d+1} \cup \partial M$ is said to admit a {\em bulk-shortcut}. 
There may be the case in which the boundary two points $p$ and $q$ are connected by a bulk null geodesic curve. 
However, if such a bulk null curve contains a pair of conjugate points, it can be deformed to a bulk timelike curve from $p$ to $q$, 
implying the existence of a bulk-shortcut.  
If there is no such a bulk-shortcut, then the achronal null geodesic segment $\gamma$ in $\partial M$ is the {\em fastest causal curve from $p$ to $q$}. 
If a bulk-shortcut exists, then it implies that a causality violation occurs in boundary field theories and therefore 
that the AdS/CFT duality would not work properly in such a bulk-boundary system. 
The {\em no-bulk-shortcut condition} is the claim that there is no bulk shortcut in the bulk-boundary system under consideration, 
and this is shown to be the case \cite{Gao:2000ga} when the bulk spacetime satisfies certain reasonabl conditions such as the ANEC.

In Ref.~\cite{IIM:2020}, we have applied the no-bulk-shortcut property above and derived some restriction to the weighted average of 
the null energy for the renormalized stress-energy tensor for boundary conformal fields. 
For convenience we provide the holographic stress-energy formulas of \cite{deHaro:2000vlm} 
here for the two and four-dimension cases. Hereafter we denote (a part of) the conformal boundary $\partial M$ by $(M_d, g_{\mu \nu})$
on which dual field theories reside.

In two dimensions, the holographic stress-energy tensor on $(M_2, g_{(0)\mu \nu})$ is given in terms of the FG expansion coefficients~(\ref{FG_coordinate}) by 
\begin{equation}
\label{se:d2}
 \braket{T_{\mu\nu}}= \frac{2 \ell}{16 \pi G} \left(g_{(2)\mu \nu} - g_{(0)\mu \nu}  {\rm Tr} \left( g_{(2)} \right) \right) \,. 
\end{equation}
Since the second term is proportional to the boundary metric $ g_{(0)} = ds^2_{\partial}$, for any null vector field $l^\mu$ 
the corresponding null energy is simply given by the contraction of $l^\mu$ with the first term $g_{(2)\mu \nu} l^\mu l^\nu$.

In four-dimensions $(M_4, g_{(0)\mu \nu})$, the coefficient $g_{(2)\mu\nu}$ is expressed by the Ricci curvature tensor of the boundary metric $g_{(0)\mu\nu}$ in the 
FG expansion~(\ref{FG_coordinate}), and the other subleading terms $g_{(4)\mu\nu}$, $h_{(4)\mu\nu}$ are 
expressed by $g_{(2)\mu\nu}$ as 
\begin{align}
\label{coefficient_FG}
& g_{(2)\mu\nu}=-\frac{1}{2}\left(R_{\mu\nu}-\frac{R}{6}g_{(0)\mu\nu}\right), \nonumber \\
& g_{(4)\mu\nu}=t_{\mu\nu}+\frac{1}{8}g_{(0)\mu\nu}\left[(\mbox{Tr} g_{(2)})^2-\mbox{Tr}(g^2_{(2)})   \right]
+\frac{1}{2}g_{(2)\mu\alpha}{g_{(0)}}^{\alpha\beta}g_{(2)\beta\nu}-\frac{1}{4}g_{(2)\mu\nu}\mbox{Tr}(g_{(2)}), 
\nonumber \\
& h_{(4)\mu\nu}=\frac{1}{2}g_{(2)\mu\alpha}{g_{(0)}}^{\alpha\beta}g_{(2)\beta\nu}
-\frac{1}{8}g_{(0)\mu\nu}\mbox{Tr}(g_{(2)}^2),  
\end{align}
where the boundary stress-energy tensor $\braket{T_{\mu\nu}}$ is related to  
the bulk tensor $t_{\mu\nu}$ in (\ref{coefficient_FG}) via the AdS/CFT duality~\cite{deHaro:2000vlm} as 
\begin{align}
\label{def:t_munu_d=4}
\braket{T_{\mu\nu}}=\frac{4}{16\pi G}t_{\mu\nu}.
\end{align} 
Here, the indices are raised and lowered by the conformal boundary metric $g_{(0)\mu\nu}$.

\section{ANEC in $2$-dimensional boundary spacetime} 
\label{sec:d=2} 

In this section we derive an ANEC for field theories on two-dimensional spacetime $(M_2,g_{\mu \nu})$, which describes either 
the spatially compact universe $R^1\times S^1$ or the spatially non-compact spacetime $R^1\times R^1$. 
As noted above, we assume that $M_2$ be realized as (a part of) the conformal boundary $\partial M$ of 
a $3$-dimensional asymptotically AdS vacuum bulk spacetime $(M_3, g_{ab})$ with the curvature scale $\ell$. 
We further assume that $(M_3,g_{ab})$ allows the FG expansion (\ref{FG_coordinate}) near the conformal boundary 
so that we can apply the holographic method of \cite{IIM:2020}. 
Since any two-dimensional spacetime is conformally flat, the two-dimensional boundary metric $ds^2_\partial$ is written in the form
\be 
\label{boundary:d=2}
ds_\p^2=e^{f(t,\varphi)}(-dt^2+d\varphi^2)=-e^{f(u,v)} dudv,   
\ee
where $v=t+\varphi, u=t-\varphi$.  
In the compact universe case, $\varphi$ is, as an angular coodinate on $S^1$, within the range $0\le \varphi\le 2\pi$.

\subsection{$d=2$ spatially compact case}
\begin{figure}
 \begin{center}
  \includegraphics[width=140mm]{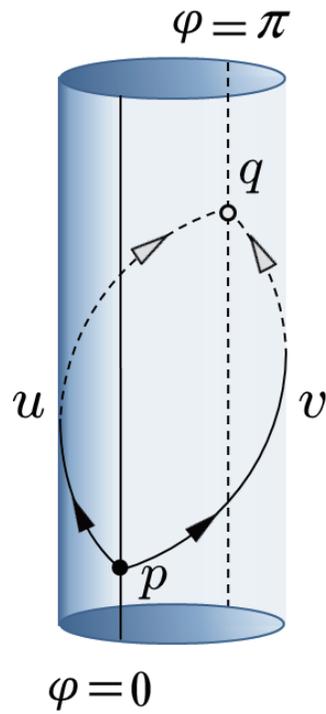}
  \caption{Null rays on $R^1\times S^1$ cylinder. Both null rays from a point $p$ at $\varphi=0$ meet up round 
  the back of the cylinder on a point $q$ at $\varphi=\pi$. The null geodesic segment with the tangent vector $\partial _v$ can only be 
  achronal between $p$ and $q$. 
  }
\label{fig1}
 \end{center}
\end{figure}

Let us consider the causal structure of the compact universe $M_2= R^1\times S^1$. As shown in Fig.~\ref{fig1}, the null rays from a point $p \in M_2$ meet up 
round the back of the cylinder on a point $q \in M_2$ at $\varphi=\pi$. Each null segment on $M_2$ connecting the two points $q$ and $p$ is achronal only when 
$\Delta \varphi\le \pi$, where $\Delta \varphi$ is defined as the coordinate length between $q$ and $p$. 
As a consequence of the no-bulk-shortcut property~\cite{Gao:2000ga}, we can establish the following theorem; 

\noindent 
{\it Theorem 1}\\ 
We assume that there is a holographic bulk dual to $M_2=R\times S^1$ and the AdS/CFT duality holds: In particular, 
FG expansion~(\ref{FG_coordinate}) and the holographic stress-energy formula~(\ref{se:d2}) apply. Furthermore assume that the
no-bulk-shortcut principle holds.
We choose the null coordinate $v$ along a null geodesic $\gamma$~(with $u=0$) on the boundary $M_2$ 
as an affine parameter and the metric function $f$ is set to $f(0,v)=0$. Then, let us consider a scalar field $\eta$ on $M_2$ 
which is not identically zero on $\gamma$ and satisfies, along $\gamma$, the following equation with the initial value $\eta(v_p)=0$ at $p$:   
\be 
\label{Jacobi:d=2}
\ddot{\eta}(v)= \frac{12 \pi}{c} \braket{T_{\mu\nu}} l^\mu l^\nu \eta(v), 
\ee
where $c$ is the central charge~\cite{BrownHenneaux} and the dot denotes the derivative with respect to $v$ and 
$\braket{T_{\mu\nu}}$ is the boundary stress-energy tensor. 
Then, there is no point $r$ on $\gamma$ between $p$ and $q$ where $\eta$ vanishes, i.e., there is no coordinate value 
$v_r \in (v_p, v_q)$ for which $\eta(v_r)=0$.  
\medskip 

We prove the theorem 1 by the holographic method below. 

\noindent 
{\it Proof}. 
Since the causal structure is invariant under the conformal transformation, we can proceed in the rescaled 
bulk spacetime $(\hat{M}_3, \,\hat{g}_{ab})$, where we define the $3$-dimensional metric $\hat{g}_{ab}=z^2g_{ab}$ so that it satisfies the FG expansion~(\ref{FG_coordinate}) with the boundary metric $g_{(0)\mu \nu}$ at $z=0$ being (\ref{boundary:d=2}). We can also extend the coordinates $(u,v)$ into $\hat{M}_3$. 
Accordingly we can view the null geodesic $\gamma$ defined on $M_2$ as a null geodesic curve $\hat{\gamma}$ embedded in $\hat{M}_3$. 
Now consider the bulk Jacobi field with respect to $\hat{\gamma}$. In the FG coordinates, 
the magunitude $\hat \eta$ of the bulk Jacobi field along the covector $(dz)_a$ obeys the following equation of motion 
\begin{equation}
\label{Rzvzveq}
 \ddot{\hat \eta} = - \hat{R}^z{}_{v z v} \,{\hat \eta}=-\hat{R}_{vv} \,{\hat \eta} \,, 
\end{equation} 
at $z=0$ 
where $\hat{R}^z{}_{vzv}$ is the corresponding component of the Riemann tensor of $({\hat M}_3, \hat{g}_{ab})$.  
By extrapolating the Eq.~\eqref{Rzvzveq} near the boundary $z \to 0$, the curvature component in the right-hand side can simply be replaced with 
$-g_{(2)\mu\nu}l^\mu l^\nu$ and using the relationship $c = 3 \ell /2 G$, we can in fact 
identify $\hat \eta$ with the boundary scalar $\eta$ using the AdS/CFT dictionary and Eq.~\eqref{Rzvzveq} reduces to Eq.~\eqref{Jacobi:d=2}. 
This implies that if there were a solution of Eq.~(\ref{Jacobi:d=2}) with $\eta(v_r)=0$ at $r$ on $\gamma \in M_2$ between $(p, q)$, 
then $r$ is conjugate to $p$ along $\gamma$ viewed as the null geodesic curve in $\hat{M}_3$. It immediately follows from the standard argument 
that there were a bulk timelike curve from $p$ to $q$. This contradicts the assumption that the null geodesic segment $\gamma$ from $p$ to $q$ is achronal and the no-bulk-shortcut condition. $\Box$

From Eq.~(\ref{Jacobi:d=2}), it is clear that if the null-null component of the boundary stress-energy tensor,  
$\braket{T_{\mu\nu}}l^\mu l^\nu$ is sufficiently negative, there is a pair of conjugate points along the null geodesic segment 
with the coordinate length less than $\Delta \phi=\pi$. As shown later, the vacuum global AdS spacetime gives the critical value. 

\subsection{$d=2$ spatially non-compact case}

We turn to the case when $M_2$ is the non-compact universe $R^1\times R^1$.  
Suppose there is a complete achronal null geodesic $\gamma$ with its tangent $l=\p_v$, which can be extended arbitrary in the future and 
past directions. In this case, we can establish the following theorem: \\
\\
{\it Theorem 2} \\
Suppose that the integral of the null energy $\braket{T_{\mu\nu}}l^\mu l^\nu$ over complete $\gamma$ converges 
to a finite value;  
\be 
\int^\infty_{-\infty}\braket{T_{\mu\nu}}l^\mu l^\nu dv=\kappa \,. 
\ee
Then, $\kappa\ge 0$ and the equality holds only if the null energy $\braket{T_{\mu\nu}}l^\mu l^\nu$ vanishes along entire $\gamma$. 

\medskip 
Again we will assume that there is a holographic bulk dual and AdS/CFT duality holds and also that no-bulk-shortcut principle holds. 
Under these assumptions we will prove the theorem 2 from the bulk. 
As in the compact case, let us consider the rescaled bulk $(\hat{M}_3, \,\hat{g}_{ab})$ and the bulk null geodesic 
congruence of the null line $\gamma$ originally defined on the conformal boundary. By the same argument in Theorem~1, the magnitude of 
the bulk Jacobi field is identified with $\eta$, which obeys Eq.~(\ref{Jacobi:d=2}). Furthermore, eq.~(\ref{Jacobi:d=2}) can be 
transformed to Raychaudhuri type equation with no shear:
\be
\label{Raychaudhuri} 
\dot{\theta}=-\frac{1}{2}\theta^2+\frac{24\pi}{c}\braket{T_{\mu\nu}}l^\mu l^\nu \,, \qquad \theta:= 2 \frac{\dot{\eta}}{\eta} \,.   
\ee 
Since $-24\pi\braket{T_{\mu\nu}}l^\mu l^\nu/c$ is identified with the null-null component of the bulk Ricci tensor 
$\hat{R}_{\mu\nu}l^\mu l^\nu$ for the rescaled bulk $(\hat{M}_3, \,\hat{g}_{ab})$, 
one can apply the focusing theorem (Theorem 2 in~\cite{Borde1987}) to show the existence of a pair of conjugate 
points of Eq.~(\ref{Raychaudhuri}). According to 
the focusing theorem, one can show that there is a pair of conjugate points along $\gamma$, provided that 
the average of $\hat{R}_{\mu\nu}l^\mu l^\nu$ is not negative and $\hat{R}_{\mu\nu}l^\mu l^\nu$ is not identically 
zero~\footnote{Even though the generic condition is not satisfied, one can still apply the focusing theorem to 
Eq.~(\ref{Raychaudhuri}) since in the present case $\hat{R}_{\mu\nu}l^\mu l^\nu$ is not identically zero.}. 

\medskip 

\noindent 
{\it Proof}. Suppose that $\kappa\le 0$ and $\braket{T}_{\mu\nu}l^\mu l^\nu$ is not identically zero. Then, by the focusing 
theorem \cite{Borde1987}, there would be a point $r$ conjugate to $p$ along $\gamma$. Then, there were 
a timelike curve from $p$ to $q>r$. This contradicts the assumption that the null geodesic $\gamma$ is achronal.  $\Box$ 

 \subsection{Examples}
To clarify the statement of the theorem 1, let us consider the vacuum bulk solution with the metric 
\begin{align}
\label{BTZ} 
& ds^2=-F(r)dt^2+\frac{dr^2}{F(r)}+r^2\left(\frac{J}{2r^2} dt+d\varphi \right)^2, \nonumber \\
& F(r)=-m+r^2+\frac{J^2}{4r^2},  
\end{align}
where $m$ and $J$ are the parameters corresponding respectively to the mass and the angular momentum. When $m>0$ and $m\ge |J|$, it corresponds to 
the BTZ black hole~\cite{BTZ}. The conformal boundary metric is written by Eq.~(\ref{boundary:d=2}) with $f=0$ and, near the boundary, 
the bulk metric can be transformed into the FG coordinate~(\ref{FG_coordinate}) by 
\be
\label{FG-transform} 
r=\frac{1}{z}+\frac{m z}{4}+\cdots. 
\ee 
According to Ref.~\cite{deHaro:2000vlm}, the tensor $t_{\mu \nu}$ in Eq.~(\ref{def:t_mu_nu_d=2}) proportional to 
the stress-energy tensor on the boundary theory can be read off from 
\be 
g_{(2)\mu\nu}=\frac{1}{2}(Rg_{(0)\mu\nu}+t_{\mu\nu}), 
\ee
where $R$ is the Ricci scalar curvature on the boundary spacetime and 
the tensor $t_{\mu\nu}$ is defined as 
\be 
\label{def:t_mu_nu_d=2}
t_{\mu\nu}:= \frac{24 \pi}{c} \braket{T_{\mu\nu}}. 
\ee
Then, each component of $t_{\mu\nu}$ becomes 
\be
\label{d=2:stress-energy} 
t_{tt}=t_{\varphi\varphi}=m, \qquad t_{t\varphi}=-J
\ee
and the null-null components along the null geodesics $l=\p_v$ and $k=\p_u$ are thus 
\be 
t_{\mu\nu}l^\mu l^\nu=\frac{1}{2}(m-J), \qquad t_{\mu\nu}k^\mu k^\nu=\frac{1}{2}(m+J). 
\ee
Without loss of genericity, suppose $J\ge 0$ and consider the Jacobi equation~(\ref{Jacobi:d=2}) along $l^\mu$. 
If $m-J\ge 0$, there is no non-trivial solution $\eta$ which has two zeros. If $m-J<0$, the solution 
\be 
\eta=\epsilon\sin\left(\frac{\sqrt{J-m}}{2}v+\delta \right)
\ee
has a pair of conjugate points with the coordinate length~(note that $\Delta v=2\Delta \varphi$ along $u=0$)
\be 
\Delta \varphi=\frac{\pi}{\sqrt{J-m}}. 
\ee 
Theorem 1 asserts that $\Delta \varphi\ge \pi$ and hence  
\be 
J-m\le 1 \,\, \Longleftrightarrow \,\, m\ge J-1. 
\ee
Therefore, the minimum value of the mass parameter $m$ is obtained at $J=0$ by $m=-1$, which corresponds to the global AdS spacetime. 
Theorem 1 also applies to some inhomogeneous bulk spacetime with, e.g., a perfect fluid star at the center.   
In general, $t_{\mu\nu}l^\mu l^\nu$ can be a function of $u$ and $v$, and it can be negative in some region. In such a 
case, if the coordinate length $\Delta \varphi$ between the two conjugate points is less than $\pi$, the bulk spacetime 
violates the no-bulk-shortcut property, yielding the pathological bulk spacetime. Thus, Theorem 1 restricts the extent of the 
possible local violation of the null energy condition.  

As an example of the spatially non-compact spacetime, let us consider a $d=2$ black hole spacetime with the metric 
\begin{align} 
ds^2=-F(r)dt^2+\frac{dr^2}{F(r)}, \nonumber 
\end{align}
where we assume $F(r)>0~(r>r_0)$, $F(r_0)=0$, and $F(r)<0~(r<r_0)$. 
It is straightforward to find the coordinate transformation in which this metric takes the double-null from~(\ref{boundary:d=2}). 
The event horizon at $r=r_0$ is the bifurcate Killing horizon and its null geodesic generator is complete and achronal. 
Theorem~2 states that the average of the null-null component of the boundary stress-energy tensor cannot be negative. In particular,  if it 
is zero, it should be identically zero. This implies that a negative null energy locally created by quantum effects can be compensated by 
larger amount of a positive null energy on any achronal null line. 
 
\section{The weighted ANEC in d=4 spatially compact universe}\label{sec:d=4}
Although its main focus was on the odd-dimension case, Ref.~\cite{IIM:2020} also briefly discussed the $4$-dimension case and derived 
the ANEC with a weight function for the $4$-dimensional static Einstein universe with compact spatial section.  
In this section, we extend the result to a class of time-dependent universe with compact spatial section, and 
show that the averaged null energy $t_{\mu\nu}l^\mu l^\nu$ with an appropriate weight function is bounded from below by the Ricci curvature tensor, and the expansions of the null vectors.   

As a boundary spacetime $M_4$, we consider the following metric 
\begin{align}
g_{\mu\nu}dx^\mu dx^\nu 
=e^{f(t,\rho)}(-dt^2+d\rho^2)+r^2(t,\rho)(d\theta^2+\sin^2\theta d\varphi^2) \,, 
\end{align}  
where the topology of the $t=const.$ hypersurface is $S^3$, and the $\rho=const.$ subspace is a two-dimensional sphere.  
For convenience, we also introduce the null-coordinates as $v=t+\rho, u=t-\rho$ 
so that the metric becomes 
\begin{align}
\label{boundary:d=4} 
g_{\mu\nu}dx^\mu dx^\nu = -e^{f(u,v)}dudv+r^2(u,v)(d\theta^2+\sin^2\theta d\varphi^2) \, .
\end{align}  
One considers a boundary null geodesic segment $\gamma$ with tangent vector $l=\p_v$ along $u=0$ null hypersurface from 
the south pole $p$~($v=v_-$) to the north pole $q$~($v=v_+$), where $r(0,v_-)=r(0,v_+)=0$. 
By a suitable coordinate transformation, one can always take
\be 
\label{condition:f}
f(0,\,v)=0, 
\ee
so that $v$ is the affine parameter. 
The Ricci tensor and the scalar curvature are given by 
\begin{align}
\label{dim_4_curvature}
& R_{vv}=\frac{2\dot{f}\dot{r}-2\ddot{r}}{r}, \qquad R_{uv}=-\dot{f'}-\frac{2\dot{r'}}{r}, \nonumber \\
& R_{\theta\theta}=\frac{R_{\varphi\varphi}}{\sin^2\theta}=1+4e^{-f}(\dot{r}r'+r\dot{r'}), \nonumber \\
& R=\frac{2e^{-f}}{r^2}\left[e^f+4\dot{r}r'+2r^2\dot{f'}+8r\dot{r'} \right], 
\end{align}
where the dot and the prime represent the derivative with respect to $v$ and $u$, respectively.  

\subsection{Derivation of the weighted ANEC}

Let us consider an achronal boundary null geodesic segment $\gamma \in M_4$, which connects the two points; 
the south pole $p$ and north pole $q$ on the boundary $M_4$. Now, near $\gamma \in M_4$, we also consider a bulk causal curve 
$\lambda$ in the rescaled manifold $(\hat{M}_5,\,\hat{g}_{ab})$ which has two endpoints at $p$ and $q$ on the boundary $M_4$.  
The tangent vector $K^a$ to $\lambda$ is written by 
\begin{align} 
& K^a=\left(\frac{dz}{dv}, \frac{du}{dv}, 1, {\bm 0}  \right), \nonumber \\
& z=\epsilon z_1+\epsilon^2z_2+\cdots, \nonumber \\
& \frac{du}{dv}=\epsilon^2\frac{du_2}{dv}+\epsilon^3\frac{du_3}{dv}+\epsilon^4\frac{du_4}{dv}
+\epsilon^4\ln \epsilon^2\,\frac{d\xi_4}{dv}+\cdots., 
\end{align}
where $\epsilon$ is an arbitrary small parameter and $z$ satisfies the boundary condition 
\begin{align}
\label{bc:z}
z(v_-)=z(v_+)=0. 
\end{align}
Due to the existence of the logarithmic term~(\ref{FG_coordinate}), one needs to 
consider the last logarithmic term in the third line, as shown below.  
 
Expanding $\hat{g}_{ab}K^a K^b\le 0$ as a series in $\epsilon$, one obtains 
\begin{align}
\label{Causal_exp_2}
g_{ab}K^aK^b 
& = \epsilon^2\left(-\frac{du_2}{dv}+\dot{z}_1^2+z_1^2\,g_{(2)vv}(0,v)\right)   \nonumber \\
&\quad +\epsilon^3\left(2\dot{z}_1\dot{z}_2+2z_1z_2\,g_{(2)vv}(0,v)-\frac{du_3}{dv}  \right) \nonumber \\
&\quad +\epsilon^4\Biggl(\dot{z}_2^2+z_2^2\,g_{(2)vv}(0,v)+2\dot{z}_1\dot{z}_3 \nonumber \\
&\qquad \qquad +2z_1z_3\,g_{(2)vv}(0,v) +2z_1^4\ln z_1\,h_{vv}(0,v) \nonumber \\
&\qquad \qquad +2z_1^2\,g_{(2)uv}(0,v)\frac{du_2}{dv}-\frac{du_4}{dv}+z_1^4\,g_{(4)vv}(0,v) \nonumber \\
&\qquad \qquad +z_1^2\p_u(g_{(2)vv})(0,v)u_2-(\p_uf)(0,v)u_2\frac{du_2}{dv} \Biggr) \nonumber \\
&\quad +2\epsilon^4\ln (\epsilon)\left(z_1^4\,h_{vv}(0,v)-\frac{d\xi_4}{dv}\right)+\cdots \nonumber \\
& \le 0.  
\end{align}
Note that the functions $g_{(2)vv}(u,v)$ and $f(u,v)$ are expanded around $u=0$ as $\zeta(u,v)=\zeta(0,v)+
\epsilon^2 \p_u(\zeta(0,v))u_2+\cdots$. 

At the leading order, $O(\epsilon^2)$, by integrating the above equation from $v=v_-$ to $v=v_+$, one obtains 
\begin{align}
\label{delta_u_2}
\Delta u\ge \epsilon^2\int^{v_+}_{v_-}(\dot{z}_1^2+z_1^2g_{(2)vv}(0,v))dv,  
\end{align}
where $\Delta u$ is the coordinate distance between $v_-$ and $v_+$, and the equality holds for the null curve. 
Applying the variational principle to the r.~h.~s. of Eq.~(\ref{delta_u_2}), we obtain the 
equation~\cite{IIM:2020}
\begin{align}
\label{z_1:eq}
\ddot{z}_1=g_{(2)vv}(0,v)z_1=\frac{\ddot{r}(0,v)}{r(0,v)}z_1.  
\end{align} 
The solution that satisfies the condition~(\ref{bc:z}) is given by 
\begin{align}
\label{z_1:sol}
z_1=r(0,v), \qquad z_1(v_-)=z_1(v_+)=0. 
\end{align}
The substitution of (\ref{z_1:eq}) into Eq.~(\ref{delta_u_2}) yields 
\begin{align}
\Delta u_2=0
\end{align}
for the bulk null curve $\lambda$. Here, note that $u_2$ is rewritten by $z_1$ as 
\begin{align}
\label{u_2:sol}
u_2=z_1\dot{z}_1. 
\end{align}
Integrating (\ref{Causal_exp_2}) by parts at $O(\epsilon^3)$, 
one can also show that 
\begin{align}
\Delta u_3=0
\end{align}
for the bulk null curve $\lambda$ satisfying the boundary condition~(\ref{bc:z}).   
 
At $O(\epsilon^4\ln (\epsilon))$ in Eq.~(\ref{Causal_exp_2}), one obtains 
\begin{align}
\label{Delta_xi}
\Delta\xi_4
& \ge \int^{v_+}_{v_-}z_1^4\,h_{vv}(0,v)dv  \nonumber \\
& =\int^{v_+}_{v_-}z_1^4\Biggl[ -r'\left\{\frac{2\dot{r}\ddot{r}+r\dddot{r}}{6r^3} \right\}-\frac{\dot{r'}\ddot{r}}{6r^2}
    +\frac{\ddot{r'}\dot{r}}{2r^2}+\frac{\dddot{r'}}{6r} \nonumber \\
 &\qquad -\frac{\dot{r}^2}{2r^2}\dot{f'}-\frac{\ddot{r}}{6r}\dot{f'}-\frac{\dot{r}}{2r}\ddot{f'}-\frac{\dddot{f'}}{12}\Biggr]dv. 
\end{align}
By using integration by parts and Eq.~(\ref{z_1:sol}), one can show that the r.~h.~s. of Eq.~(\ref{Delta_xi}) is zero. 
Thus, for the null curve $\lambda$, there is no time delay at this order, i.~e.~, $\Delta \xi_4=0$, independent of the time dependence 
of the boundary metric~(\ref{boundary:d=4}). Therefore, the time delay between the bulk null curve $\lambda$ and the boundary null geodesic $\gamma$ is caused 
by $O(\epsilon^4)$ in Eq.~(\ref{Causal_exp_2}). 
 
Just like the $O(\epsilon^2)$ case, $\Delta u_4$ is minimized by $z_2$ satisfying 
\begin{align}
\label{z_2:eq}
\ddot{z}_2=g_{(2)vv}(0,v)z_2=\frac{\ddot{r}(0,v)}{r(0,v)}z_2  
\end{align} 
whose solution is given by 
\begin{align}
\label{z_2:sol}
z_2=\alpha r(0,v) 
\end{align}
with a constant $\alpha$.  Substituting Eqs.~(\ref{coefficient_FG}), (\ref{z_2:sol}), and (\ref{u_2:sol}) into 
Eq.~(\ref{Causal_exp_2}) and integrating by parts, one obtains 
\begin{align}
\label{Delta_u_4} 
\Delta u_4
& \ge \int^{v_+}_{v_-} z_1^4\,t_{vv}(0,v)dv+2\int^{v_+}_{v_-} z_1^4\ln z_1\,h_{(4)vv}(0,v)dv \nonumber \\
&\qquad -\frac{1}{4}\int^{v_+}_{v_-}g_{(2)vv}(0,v)\mbox{Tr}\{g_{(2)}(0,v)\}z_1^4dv
-\frac{1}{4}\int^{v_+}_{v_-}z_1^4\,\p_v(\p_u(g_{(2)vv})(0,v))dv \nonumber \\
& \qquad +\int^{v_+}_{v_-}z_1^2\dot{z}_1^2\left[\frac{1}{4}\left(\frac{1}{r^2}+\theta_+\theta_-\right)
-\frac{5R_{\theta\theta}}{12r^2}-\frac{7}{6}R_{uv} \right]dv \nonumber \\
& \ge 0, 
\end{align}
where $\theta_\pm$ are the expansions along the null vector $\p_v$ and $\p_u$ defined by 
\begin{align}
\theta_+:=\frac{2\dot{r}(0,v)}{r(0,v)}, \qquad \theta_-:=\frac{2r'(0,v)}{r(0,v)}. 
\end{align} 
Here, the equality in the first line holds for the bulk null curve $\lambda$ and 
the inequality in the last line comes from the no bulk-shortcut principle. 

In general, the r.~h.~s. of Eq.~(\ref{Delta_u_4}) includes fourth derivatives of the boundary metric functions $f$ and $r$. As shown 
in the Appendix, by performing integration by parts, the inequality~(\ref{Delta_u_4}) can be expressed by expansions $\theta_{\pm}$ 
and the curvature on the boundary spacetime (\ref{boundary:d=4}) as
\begin{align}
\label{energy_inequality}
& \int^{v_+}_{v_-} \eta^4\,t_{vv}(0,v)dv \nonumber \\
&\ge\frac{1}{12}\int^{v_+}_{v_-}\eta^4\Biggl[\theta_+^2\left(R_{uv}+\frac{R_{\theta\theta}}{r^2} \right)
-\frac{\theta_+^2}{2} \cdot \mu(r)   
                     +\frac{1}{2}\theta_+\theta_-R_{vv} \nonumber \\
&\qquad \qquad +\left(\frac{3}{2r^2}-\frac{R_{\theta\theta}}{r^2}-2R_{uv}\right)R_{vv} \Biggr]dv 
\end{align} 
with the help of Eqs.~(\ref{formula_1}), (\ref{formula_2}), and ~(\ref{formula_3}), where $\mu$ defined by  
\begin{align}
\mu (r)  := \left(\frac{1}{r^2}+\theta_+\theta_- \right) \,, 
\end{align}
is the quasi-local mass density, i.e., whose integral over two-sphere provides the quasi-local gravitational mass~\cite{Hayward:1994bu}.   
Here, $\eta$ is the Jacobi field of the null geodesic congruence of the boundary spacetime~(\ref{boundary:d=4}) and 
it is proportional to $z_1$, just like the case~\cite{IIM:2020}. 
This is the averaged null energy condition in $d=4$, weighted 
by the Jacobi field $\eta$. 
Since the second term of the r.~h.~s. of (\ref{energy_inequality}) is the quasi-local 
mass density $\mu$ of the boundary spacetime with weight function $\eta^4\theta_+^2$, 
the averaged null energy is bounded by the local mass density when the Ricci curvature is small enough compared with the 
expansions.        

As discussed in Sec.~\ref{sec:ex:negative}, the equality should hold when the boundary state becomes the ground state. 
The r.~h.~s. of the inequality~(\ref{energy_inequality}) gives the weighted average of the null energy on the ground state. 
In particular, when the boundary spacetime includes horizons, or wormhole throat with zero expansion, i.~e.~, $\theta_+=0$, 
the integrand of the r.~h.~s. of the inequality~(\ref{energy_inequality}) reduces to a simple form
\begin{align} 
& \theta_+^2\left(R_{uv}+\frac{R_{\theta\theta}}{r^2} \right)
-\frac{\theta_+^2}{2} \cdot \mu (r)
+\frac{1}{2}\theta_+\theta_-R_{vv} \nonumber \\
&+\left(\frac{3}{2r^2}-\frac{R_{\theta\theta}}{r^2}-2R_{uv}\right)R_{vv}
=-\left(\frac{1}{r^2}+4\dot{f'} \right)\frac{\ddot{r}}{r},  
\end{align}
with the help of Eqs.~(\ref{dim_4_curvature}) and the condition~(\ref{condition:f}). Therefore if we consider, for example, 
the boundary spacetime with a wormhole throat at $v=v_0$, the radius of the throat takes its minimum there, i.~e.~, 
\begin{align}
\ddot{r}(v_0)>0. 
\end{align}  
This suggests that the averaged null energy of the boundary theory becomes negative when the throat radius, $r(v_0)$, is  
small enough on the ground state. An example of such a wormhole geometry will be given in Sec.~\ref{sec:ex:negative}.   

\subsection{Schwarzschild-AdS bulk and boundary ANEC}
Let us examine the the averaged null energy condition~(\ref{energy_inequality}) when our $5$-dimensional bulk spacetime $M_5$ is 
given by the Schwarzschild-AdS metric  
\begin{align}
\label{5d_Sch}
ds^2=-\left(r^2+1-\frac{M}{r^2}  \right)dt^2+\left(r^2+1-\frac{M}{r^2}  \right)^{-1}dr^2+r^2(d\rho^2+\sin^2\rho d\Omega^2), 
\end{align} 
where $d\Omega^2$ is the metric of the unit two-dimensional sphere. The conformal boundary $M_4$ is the static Einstein 
universe whose metric is expressed by 
\begin{align} 
\label{boundary_metric_Sch}
 ds^2_\p=-dudv+\sin^2\left(\frac{v-u}{2}  \right)(d\theta^2+\sin^2\theta d\varphi^2),  \nonumber 
 \end{align}
where $\rho=({v-u})/{2}, t=({v+u})/{2}$. 
The FG coordinate~(\ref{FG_coordinate}) is obtained by the following coordinate transformation near the boundary: 
\begin{align}
r(z)=\frac{1}{z}-\frac{1}{4}z+\frac{M}{8}z^3+O(z^4). 
\end{align}
The metric $g_{\mu\nu}$ in the above coordinate system can be expanded as 
\begin{align}
\label{coefficient_g_Sch}
& g_{vv}=g_{uu}=-\frac{z^2}{4}+\frac{M}{4}z^4+\cdots, \nonumber \\
& g_{uv}=-\frac{1}{2}+\frac{1}{8}\left(M-\frac{1}{4}  \right)z^4+\cdots, \nonumber \\
& g_{\theta\theta}=\frac{g_{\varphi\varphi}}{\sin^2\theta}
=\sin^2\chi-\frac{1}{2}\sin^2\chi\,z^2+\frac{1}{16}\left(1+4M  \right)z^4+\cdots. 
\end{align}
Substituting Eq.~(\ref{coefficient_g_Sch}) into Eq.~(\ref{coefficient_FG}), we obtain the null-null component of the 
tensor $t_{\mu\nu}$ in Eq.~(\ref{def:t_munu_d=4}); 
\begin{align}
\label{null_energy_Sch}
t_{vv}=\frac{1}{16}(4M+1). 
\end{align}
On the other hand, the Ricci curvature of the boundary metric~(\ref{boundary_metric_Sch}) is given by 
\begin{align}
\label{Ricci_curvature_Sch}
R_{vv}=\frac{1}{2}, \quad R_{uv}=-\frac{1}{2}, \quad R_{\theta\theta}=2\sin^2\chi, \quad 
R_{\varphi\varphi}=2\sin^2\chi\sin^2\theta. 
\end{align}
Substitution of Eqs.~(\ref{null_energy_Sch}) and (\ref{Ricci_curvature_Sch}) into Eq.~(\ref{energy_inequality}) yields 
\begin{align}
\frac{3(4M+1)\pi}{64}\alpha^4 
& \ge  
\frac{1}{12}\int^{2\pi}_{0}\eta^4\Biggl[\theta_+^2\left(R_{uv}+\frac{R_{\theta\theta}}{r^2} \right)
-\frac{\theta_+^2}{2} \cdot \mu 
 +\frac{1}{2}\theta_+\theta_-R_{vv} \nonumber \\ 
& \qquad \qquad +\frac{3}{2r^2}R_{vv}-\frac{R_{\theta\theta}}{r^2}R_{vv}-2R_{vv}R_{uv} \Biggr]dv
\nonumber \\
& =\frac{3\pi}{64}\alpha^4,  
\end{align}
where $v_-=0$ and $v_+=2\pi$, and $\eta=\alpha z_1=\alpha r(0,v)$ for a constant $\alpha$. This inequality means 
that the mass parameter must be non-negative; 
\be
M\ge 0. 
\ee
This example suggests that the no bulk-shortcut principle is connected with the positive mass theorem in asymptotically 
anti de Sitter spacetime. This is because the Schwarzschld-AdS spacetime with negative mass $M~(<0)$ has a naked 
singularity on the bulk, and hence, it is predicted by the theorem~\cite{IMM:2019}, which prohibits the appearance 
of naked singularities.    

\section{Minimum of the null energy in $d=4$ spatially compact universe}\label{sec:mim:energy}
The averaged null energy condition~(\ref{energy_inequality}) restricts the extent of how negative null energy appears in the 
spatially compact spacetime. However, the condition cannot tell us how and when the equality in Eq.~(\ref{energy_inequality}) holds. 
According to the AdS/CFT duality~\cite{Maldacena:1997re}, the boundary stress-energy tensor is determined not only by  
the boundary source~(conformal boundary metric) but also by the state of the boundary quantum fields. However, it appears to be 
reasonable to expect that the minimal null energy could be determined merely by the conformal boundary metric itself, provided that 
the boundary field theory has a stable ground state and also that there is no pathological behavior 
such as naked singularity or causality violating region in the dual bulk.   
The example in the previous section is a particular case in the sense that the boundary geometry~(\ref{boundary:d=4}) is 
the static Einstein universe and we would like to know whether the minimum (averaged) 
null energy is given by the r.~h.~s. of Eq.~(\ref{energy_inequality}) in a more general class of boundary spacetimes. 
In this section, we study if the equality in Eq.~(\ref{energy_inequality}) holds for a deformed static Einstein boundary universe by performing linear perturbations in the global AdS vacuum bulk.  
\subsection{The perturbed static vacuum bulk and the boundary null energy}
We consider, as our five-dimensional bulk spacetime $M_5$, the global AdS spacetime with the unit curvature length,  
\begin{align} 
\label{global_AdS_metric}  
 ds^2 =& \, g_{ab}dy^a dy^b+r(y)^2\gamma_{ij}dz^i dz^j, \nonumber \\ 
 g_{ab}dy^a dy^b :=& -(1+r^2)dt^2+\frac{dr^2}{(1+r^2)}, \nonumber \\
 \gamma_{ij}dz^i dz^j =& \, d\rho^2+\sin^2\rho(d\theta^2+\sin^2\theta d\varphi^2), 
\end{align}
where $y^a=(t,r)$ denote the static coordinates in the two-dimensional part of the global AdS metric and $z^i = (\rho, \theta, \varphi)$ 
the angle coordinates of the unit three-sphere.  Through this section, the latin indices in the range $a,b,\dots, h$ are used to denote tensors in 
the two-dimensional spacetime spanned by $y^a$ and should not be confused with the indices for tensors in the bulk. 
It is easy to obtain the FG metric~(\ref{FG_coordinate}) by the coordinate transformation  
\begin{align}
\label{FG_trans_r_z}
r=\frac{1-z^2}{2z}, \quad \tau:=\frac{t}{2}, \quad g_{ab}dy^a dy^b=\frac{1}{z^2} \left\{ dz^2 -(1+z^2)^2d\tau^2 \right\} \, .   
\end{align}
The four-dimensional boundary metric $g_{(0) \mu \nu}$ is given by the coordinates $x^\mu = (\tau, z^i) = (\tau, \rho, \theta, \varphi)$.  

By considering static metric perturbations on this background, we construct an asymptotically AdS, deformed static vacuum bulk.  
For this purpose, we follow Ref.~\cite{WaldIshibashi04} in which thorough analysis of linear perturbations on the global AdS spacetime 
has been performed.  
The relevant perturbations are of the scalar-type in the classification of \cite{WaldIshibashi04}, 
which behave as scalar fields with respect to coordinate changes in the $3$-sphere $\gamma_{ij}$.  Accordingly, the scalar-type metric perturbations can be expanded in terms of the scalar harmonics $\mS$ on the $3$-sphere that solve the equation 
\begin{align} 
\label{harmonic}
\left(D^i D_i+k(k+2)\right)\mS_k=0,  \quad k=1,2,\cdots
\end{align}
where  $D_i$ is the covariant derivative operator of the metric $\gamma_{ij}$. 
Note that $k=0$ mode corresponds to the homogeneous perturbation with respect to $\gamma_{ij}$ and 
is not relevant for the present purpose.

The solutions of the harmonic equation~(\ref{harmonic}) are given by the Jacobi polynomial as 
\begin{align}
\label{Jacobi_poly} 
& \mS_k(\rho)=P_k^{\frac{1}{2}, \frac{1}{2}}(\xi), \nonumber \\
& P_k^{\frac{1}{2}, \frac{1}{2}}(\xi)=\frac{(-1)^k}{2^k k!(1-\xi)^\frac{1}{2}(1+\xi)^\frac{1}{2}}
\frac{d^k}{d\xi^k}\{(1-\xi)^{\frac{1}{2}+k}(1+\xi)^{k+\frac{1}{2}}   \}, \nonumber \\
& P_0^{\frac{1}{2}, \frac{1}{2}}(\xi)=1, \quad P_1^{\frac{1}{2}, \frac{1}{2}}(\xi)=\frac{3}{2}\xi, \quad 
P_2^{\frac{1}{2}, \frac{1}{2}}(\xi)=\frac{5}{8}(4\xi^2-1), \cdots \,,  
\end{align}
where $\xi := \cos \rho$. 

In the scalar-type metric perturbation, the perturbed metric is generally written in the form 
\begin{align}
\label{general_form_per}
\delta ds^2=\epsilon(h_{ab}dy^a dy^b+2h_{ai}dy^adz^i+h_L\gamma_{ij}dz^i dz^j)
+\epsilon \left(D_i D_j-\frac{1}{3}\gamma_{ij}D^m D_m  \right)h_T,  
\end{align}
where $\epsilon$ is an arbitrary small parameter, and $h_{ab}$, $h_L$, and $h_T$ are expanded by $\mS_k$ as 
\begin{align}
& h_{ab}(r,\rho)=\sum_{k=1}^\infty H_{k, ab}(r) \mS_k(\rho), \quad h_L(r, \rho)=\sum_{k=1}^\infty H_{k, L}(r) \mS_k(\rho), 
\nonumber \\ 
& h_a=\sum_{k=1}^\infty H_{k, a}(r) \mS_k(\rho), \quad   h_T(r, \rho)=\sum_{k=1}^\infty H_{k, T}(r) \mS_k(\rho). 
\end{align}
The gauge-invariant variables $Z$ and $Z_{ab}$ are composed of the perturbed functions as   
\begin{align}
\label{gauge_invariant_vari}
& Z:=\frac{3}{r}\left\{H_L+\frac{k(k+2)}{3}H_T+2r(\nabla^ar)X_a     \right\}, \nonumber \\
& Z_{ab}:=r(H_{ab}+\nabla_a X_b+\nabla_bX_a)+\frac{2}{3}Zg_{ab}, 
\end{align}
where $\nabla_a$ is the covariant derivative with respect to the metric $g_{ab}$ and $X_a$ is defined by 
\begin{align}
X_a:=-H_a+\frac{1}{2}r^2\nabla_a\left(\frac{H_T}{r^2} \right). 
\end{align} 
These variables are obtained from the master variable $\Phi_s$ as 
\begin{align}
\label{Z_Phi_s_relation}
Z_{ab}=\left(\nabla_a\nabla_b-g_{ab}\right)(r^{3/2}\Phi_s), \qquad Z={Z^a}_a,    
\end{align}
and the equation for $\Phi_s$ is described in terms of a new coordinate $x$ as 
\begin{align} 
\left(\frac{\p^2}{\p x^2}+\frac{{1}/{4}}{\sin^2 x}-\frac{ {3}/{4}+k(k+2)}{\cos^2 x}  \right)\Phi_s=0, \qquad 
r=\frac{\cos x}{\sin x}. 
\end{align}

Since we are interested in the static perturbation of the global vacuum AdS spacetime, in which there is no horizon 
at the center, we need to impose the regularity condition at the center.  
Such a regular solution is given, in terms of the hypergeometric function, by 
\begin{align}
\Phi_s &=B_1(\sin x)^\frac{1}{2}(\cos x)^{\sigma+\frac{1}{2}}
F(\zeta_\sigma, \, \zeta_\sigma, \, 1+\sigma;\, \cos^2 x) \nonumber \\
&=B_1\cdot G(x)\frac{\Gamma(1+\sigma)}{\Gamma(\zeta_\sigma)^2}\sum_{k=0}^\infty 
\frac{\{(\zeta_\sigma)_k\}^2}{(k!)^2}\cdot (\sin x)^{2k} \times \nonumber \\
&{} \qquad \qquad \cdot \{2\psi(k+1)-2\psi(\zeta_\sigma+k)-\ln (\sin^2 x) \},   
\label{Hypergeometric}
\end{align}
where the parameters $\sigma, \zeta_\sigma, (\zeta)_k$ and the functions $G(x), \psi(x)$ are defined by 
\begin{align} 
&\sigma:=k+1, \qquad \zeta_\sigma=\frac{\sigma+1}{2}=\frac{k+2}{2}, \qquad 
(\zeta)_k:=\frac{\Gamma(\zeta+k)}{\Gamma(\zeta)},  \nonumber \\
& G(x)=(\cos x)^{\sigma+\frac{1}{2}}\cdot (\sin x)^\frac{1}{2}, \qquad \psi(x):=\frac{d}{dx}\log \Gamma(x) \,. 
\end{align}
\subsection{The Fefferman-Graham gauge}
In order to reconstruct the metric functions~(\ref{general_form_per}) from the master variable, one needs to fix the gauge. One may 
think of the following gauge:  
\begin{align}
\label{simplest_gauge}
H_a=H_T=0~(X_a=0). 
\end{align}
In this case, however, it turns out to be difficult to transform the metric to the Fefferman-Graham gauge~(\ref{FG_trans_r_z}) since 
$H_{zz}\neq 0$. So, our strategy is to first take the gauge~(\ref{simplest_gauge}) and transfom the perturbed metric 
to attain the gauge $H_{zz}=H_z=0$ by using gauge-freedom: 
\begin{align} 
& H_{ab}\to H_{ab}-\nabla_a\xi_b-\nabla_a \xi_b, \nonumber \\
& H_a\to H_a-\xi_a-r^2\nabla_a\left(\frac{\xi}{r^2} \right). 
\end{align}
In the static perturbation, $\xi_a dx^a=\xi_z dz$, and $\xi_z$ and $\xi$ are obtained by solving 
\begin{align} 
\label{Eq:gauge_z}
H_{zz}-2\xi_z'-\frac{2}{z}\xi_z=0
\end{align}
and 
\begin{align} 
\label{Eq:gauge_xi}
\xi_z+r(z)^2\p_z\left(\frac{\xi}{r(z)^2} \right)=0. 
\end{align}
Once $\xi$ and $\xi_z$ are obtained from the above equations, the other variables are obtained by 
\begin{align} 
\label{gauge_transformation}
& H_T\to 0-2\xi, \nonumber \\
& H_L\to H_L+\frac{2k_s^2}{3}\xi-2r(\nabla^a r)\xi_a=H_L+\frac{2k(k+2)}{3}\xi+\frac{1-z^4}{2z}\xi_z. 
\end{align}
\subsection{The stress-energy tensor in the deformed boundary metric}
For simplicity, we hereafter restrict our attention to the $k=2$~($\sigma=3$) mode in Eq.~(\ref{harmonic})~\footnote{$k=1$ mode is the 
odd function with respect to the equatorial plane~($\rho=\pi/2$), so the both sides of the inequality~(\ref{energy_inequality}) 
would be zero at the linear order in $\epsilon$. Thus, we consider the next mode $k=2$.}. 

By using the relation between $x$ and $z$, 
\begin{align}
\label{relation_z_x} 
\cos x=\frac{1-z^2}{1+z^2}, \qquad \sin x=\frac{2z}{1+z^2},  
\end{align}
$\Phi_s$ in Eq.~(\ref{Hypergeometric}) is expanded as 
\begin{align} 
\label{Phi_s_expansion}
& \Phi_s
=\epsilon\frac{(2z)^\frac{1}{2}(1-z^2)^\frac{7}{2}}{(1+z^2)^4}\sum_{i=0}^\infty \left(\frac{2z}{1+z^2}\right)^{2i}
\left\{a_i-2b_i \ln\left(\frac{2z}{1+z^2} \right) \right\}, \nonumber \\
& a_i:=\{2\psi(i+1)-2\psi(\zeta_3+i)\}b_i, \qquad b_i:=\frac{\{(\zeta_3)_i\}^2}{(i!)^2}. 
\end{align} 

By solving Eq.~(\ref{Eq:gauge_z}), we can obtain $\xi_z$ with a constant of integration. We fix the constant so that 
the $\tau\tau$-component of the boundary metric is set to $-1$, i.e~, 
\begin{align}
\label{constant_xi_z}
\lim_{z\to 0}z^2H_{\tau\tau}=0, 
\end{align}
and thus $\xi_z$ becomes 
\begin{align}
\label{sol:xi_z}
\xi_z=\frac{2b_0}{z}-\frac{2b_0}{3}(1+\ln 256+8\ln z)+\cdots. 
\end{align}
One can always take the condition~(\ref{constant_xi_z}) by using a freedom to choose a conformal factor of the 
boundary metric.  

$\xi$ is also obtained from Eq.~(\ref{Eq:gauge_xi}) with an integration of constant. One can choose the constant so that 
the south and north poles~($\rho=0$,\,$\pi$) are regular. As a result, $\xi$ is obtained as 
\begin{align} 
\label{sol:xi}
\xi=-\frac{b_0}{2}+\frac{b_0}{3}(1+\ln 16+4\ln z)z^2+\cdots. 
\end{align}
By using Eqs.~(\ref{gauge_transformation}), (\ref{sol:xi_z}), and (\ref{sol:xi}), one finally obtains the perturbed metric
\begin{align}
\label{sol:perturbed_metric} 
& H_{\tau\tau}=-\frac{40b_0}{3}-\frac{40b_0}{3}(3+\ln 256+8\ln z)z^2+\cdots, \nonumber \\
& H_L=\frac{b_0}{z^2}-\frac{16b_0}{3}+\frac{b_0}{9}(19-80\ln (2z))z^2+\cdots, \nonumber \\
& H_T=b_0-\frac{2b_0}{3}(1+4\ln 2+4\ln z)z^2+\cdots \,,   
\end{align}
and the coefficients $g_{(n)\mu\nu}$ of the Fefferman-Graham coordinate~(\ref{FG_coordinate}) as 
\begin{align} 
\label{g0_deformed_AdS}
g_{(0)\mu \nu}dx^\mu dx^\nu
& =-d\tau^2+\frac{1}{4}\left[1+\frac{5}{2}\epsilon b_0(1+2\cos 2\rho) \right]d\rho^2 \nonumber \\
& \quad +\frac{1}{4}\sin^2\rho\left[1+\frac{5}{2}\epsilon b_0(1+2\cos 2\rho)   \right](d\theta^2+\sin^2\theta d\varphi^2), 
\end{align}
\begin{align} 
\label{g2_deformed_AdS}
g_{(2)\mu\nu}dx^\mu dx^\nu 
 &= -\left[2+\frac{25b_0}{3}\epsilon(1+2\cos 2\rho) \right]d\tau^2
-\left[\frac{1}{2}+\frac{5b_0}{3}\epsilon(1+5\cos 2\rho) \right]d\rho^2 \nonumber \\
&\quad -\left[\frac{1}{2}+\frac{5b_0}{6}\epsilon(5+7\cos 2\rho) \right]\sin^2\rho(d\theta^2+\sin^2\theta d\varphi^2), 
\end{align}
\begin{align} 
\label{g4_deformed_AdS}
g_{(4)\mu\nu}dx^\mu dx^\nu 
 &=-\left[1+\frac{25b_0}{3}\epsilon(1+2\cos 2\rho)(3+\ln 256) \right]d\tau^2 \nonumber \\
 &\quad +\left[\frac{1}{4}+\frac{5b_0}{24}\epsilon\left\{1-48\ln 2+(18-32\ln 2)\cos 2\rho\right\} \right]d\rho^2 \nonumber \\
 &\quad +\left[\frac{1}{4}+\frac{5b_0}{24}\epsilon\{9-16\ln 2+(10-64\ln 2)\cos 2\rho\} \right]
 \sin^2\rho(d\theta^2+\sin^2\theta d\varphi^2),  
\end{align}
up to $O(\epsilon)$. 

Now we can examine the boundary null energy in this perturbatively deformed spacetime.  
In terms of the double-null coordinate~(\ref{boundary:d=4}) with Eq.~(\ref{condition:f}) for  
\begin{align} 
& du=d\tau-\frac{\sqrt{g}}{2}d\rho, \qquad dv=d\tau+\frac{\sqrt{g}}{2}d\rho, \nonumber \\
& g(\rho):=1+\frac{5}{2}\epsilon b_0(1+2\cos 2\rho), 
\end{align}
the null-null component of the tensor $t_{\mu\nu}$ defined in Eq.~(\ref{def:t_munu_d=4}) in the 
inequality~(\ref{energy_inequality}) is obtained as 
\begin{align}
\label{null_null_component:deformed} 
t_{vv}=1+\frac{5}{3}b_0\epsilon\,\{2+3\cos(2\rho)\}(1-\ln 256), 
\end{align}
up to $O(\epsilon)$. 
On the other hand, the curvature tensors and the expansions $\theta_\pm$ appearing on the r.~h.~s. of the 
inequality~(\ref{energy_inequality}) are calculated as 
\begin{align} 
\label{component_curvature_deformed}
& R_{uu}=2+5b_0\epsilon(1+4\cos (2\rho)), \nonumber \\
& R_{vv}=2+5b_0\epsilon(1+4\cos (2\rho)), \nonumber \\
& R_{uv}=-2-5b_0\epsilon(1+4\cos (2\rho)), \nonumber \\
& R_{\theta\theta}=\frac{R_{\varphi\varphi}}{\sin^2\theta}=\{2+5b_0\epsilon(3+5\cos (2\rho))\}\sin^2\rho, 
\end{align}
\begin{align} 
\label{expansion_deformed}
\theta_+=-\theta_-=2\cot\rho+\frac{5}{2}b_0\epsilon\{\cos(3\rho)-4\cos\rho\}\csc\rho, 
\end{align}
up to $O(\epsilon)$. 
Substituting Eqs.~(\ref{null_null_component:deformed}), (\ref{component_curvature_deformed}), and (\ref{expansion_deformed}) 
into the inequality~(\ref{energy_inequality}), one can check that the equality in (\ref{energy_inequality}) holds in the defomed static 
Einstein universe, up to $O(\epsilon)$. 
This implies that the minimum of the averaged null energy of the boundary stress-energy tensor with the 
weight function $\eta^4$ is expressed by the combination of the Ricci curvature and the expansions, as shown in 
the r.~h.~s. of the inequality~(\ref{energy_inequality}). 

\section{An example of negative averaged null energy} 
\label{sec:ex:negative}
In the previous sections, we have shown that the equality in Eq.~(\ref{energy_inequality}) holds for general deformed static vacuum 
AdS spacetime within the framework of linear perturbation. Here, assuming that the equality holds for the ground state in 
the boundary theory, we examine whether the averaged null energy becomes negative or not.    

\begin{figure}
 \begin{center}
  \includegraphics[width=70mm]{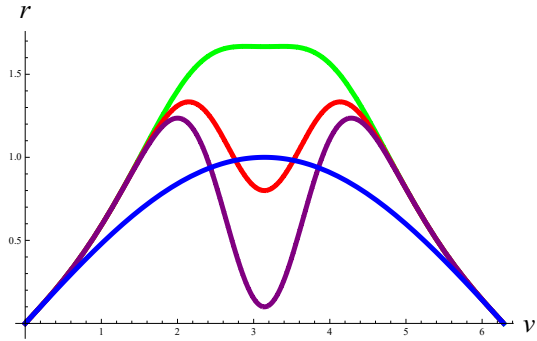}
 \caption{The function $r(v)$ is shown for various parameters of $\epsilon$ with $l=1$ and $n=6$. The dashed~(green), 
 dotted~(red), and dot-dashed~(purple) curves correspond to $\epsilon=5/3$, $4/5$, and $1/10$, respectively. The solid (blue) 
 curve is for $r=\sin(v/2)$, corresponding to the static Einstein universe. As $\epsilon$ becomes small, a wormhole throat 
 appears at $v=\pi$ and the radius of the throat becomes small.}
 \end{center}
\end{figure}
 
We consider a class of static boundary spacetimes with the metric   
\begin{align} 
\label{boundary_metric_deformed}
& ds^2_\p=-dudv+r^2(v-u)(d\theta^2+\sin^2\theta d\varphi^2),  \nonumber \\
& r(v-u)=\sin \left(\frac{v-u}{2}  \right) \cdot 
                  \left\{ 1+l\sin^2\left(\frac{v-u}{2}\right)-(1+l-\epsilon)\sin^{2n}\left(\frac{v-u}{2}\right) \right\} 
\end{align}
where $l$, $\epsilon$ are positive constants, and $n$ denotes a positive integar. As $\epsilon$ becomes smaller, with keeping $n$ a large 
value, $r$ takes a highly concave shape with the local minimum at $v-u=\pi$. The figure plots the radial function $r(v-u)$ for various values of $\epsilon$ with 
$l=1$ and $n=6$. The integral of the r.~h.~s.~of Eq.~(\ref{energy_inequality}) is analytically calculated and 
the values are 
$0.16\alpha^4$ for $\epsilon=5/3$,  $0.197 \alpha^4$ for $\epsilon=4/5$,  and $-0.995\alpha^4$ for $\epsilon=1/10$. 
This indicates that sharp concave yields large negative null energy, 
and thus, the averaged null energy can become negative.

\section{Summary}\label{sec:summary}
We have studied averaged null energy conditions~(ANEC) in $2$ and $4$ dimensional boundary theories with gravity dual. 
The basic principle we used is the no-bulk-shortcut principle, which states that there is no bulk causal curve that can 
travel faster than the boundary achronal null geodesics. In $2$-dimensional boundary spacetime, 
the null-null component of the boundary stress-energy tensor affects, via holographic argument, the behavior of boundary 
null geodesics. Namely, the achronality of a boundary null geodesic is translated to the behavior of a bulk Jacobi field. 
On the assumption of the no-bulk-shortcut principle, the bulk Jacobi field cannot behave so as to admit a pair of conjugate points 
along the achronal null geodesic line~(or segment), otherwise, there would be a bulk timelike curve which connects two boundary points 
on the achronal null geodesic, leading to a contradiction.  
For the spatially non-compact case, theorem 2 in Sec.~\ref{sec:d=2} states that the averaged null energy cannot be negative, 
agreeing with the ANEC derived in the flat spacetime~\cite{Kelly:2014mra}. 
This is applied to the geometry with black hole or cosmological horizons.  
On the other hand, for the spatially compact case such as $R^1\times S^1$ cylinder, 
the averaged null energy can become negative, but it is bounded from below, as shown in theorem 1 in Sec.~\ref{sec:d=2}.   
  
In the four-dimensional boundary spacetime case, we have derived the inequality which bounds the averaged null energy from 
below for a class of spatially compact spacetimes. The averaged null energy is bounded by the boundary geometric quantities 
such as the expansions of the boundary null geodesics and curvatures, which stem from the gravitational conformal anomalies. 
When one considers the null geodesic that goes along the Killing horizon of a black hole or passes through a wormhole throat 
with vanishing expansion, the lower bound is described by the Ricci curvature tensor. 
In particular, when the achronal null geodesic passes through a wormhole with a highly concave throat, 
the averaged null energy can become negative, due to the existence of the term $R_{\mu\nu}l^\mu l^\nu$ 
which is negative enough, as shown in Sec.~\ref{sec:ex:negative}.       
 
The ANEC~(\ref{energy_inequality}) with an appropriate weight $\eta^4$ is very similar to the conformally invariant averaged 
null energy condition~(CANEC) derived in the odd-dimensional case~\cite{IIM:2020}. 
It would be interesting to check how our inequality~(\ref{energy_inequality}) behaves under conformal transformation. 

Although the ANEC~(\ref{energy_inequality}) can be applied to non-static 
universe such as expanding cosmology, we have assumed that the boundary geometry has two-dimensional spherical 
cross-section. In this case, the boundary null geodesic congruence 
has no shear. In general, if the shear is large enough, the boundary null geodesic congruence has a pair of conjugate points within 
a small segment of the null geodesic, and then, there is a boundary timelike curve which connects two points on the null geodesic 
beyond the segment. So, beyond the small segment, the null geodesic segment is no longer achronal. 
This implies that the no-bulk-shortcut principle cannot be easily violated for the boundary spacetime with large shear. 
It would be also interesting to investigate the ANEC in such a general class of spacetimes with shear. 
  
\bigskip
\goodbreak
\centerline{\bf Acknowledgments}
\noindent
This work was supported in part by JSPS KAKENHI Grant No. 18K03619 (N.I.), 20K03938 (A.I.), 17K05451 (K.M.).

\appendix
\section{General formulas}
In this appendix, we give useful formulas to derive the null energy inequality. Integration by parts gives 
\begin{align}
\label{formula_1}
& \int^{v_+}_{v_-}z_1^4\,\p_v(\p_u(g_{(2)vv})(0,v))dv \nonumber \\
&=-4\int^{v_+}_{v_-}z_1^3\dot{z}_1\p_u(g_{(2)vv})(0,v)dv \nonumber \\
&=-4\int^{v_+}_{v_-}z_1^4\frac{\dot{r}}{r}\p_u(g_{(2)vv})(0,v)dv \nonumber \\
&=\int^{v_+}_{v_-}z_1^4\theta_+ \p_u(R_{vv})dv
=\int^{v_+}_{v_-}z_1^4\theta_+\left(\theta_+\dot{f'}-\frac{\theta_-}{2}R_{vv}-\frac{2\ddot{r'}}{r} \right)dv, 
\end{align}
where Eq.~(\ref{dim_4_curvature}) is used in the last line. The last term in the last line is expressed by the expansions $\theta_\pm$ 
and the Ricci curvature in Eqs.~(\ref{dim_4_curvature}) as 
 \begin{align}
\label{formula_2}
&-\int^{v_+}_{v_-}z_1^4\theta_+\frac{2\ddot{r'}}{r}dv \nonumber \\
&=-2\int^{v_+}_{v_-}\theta_+r^3\ddot{r'}dv \nonumber \\
&=2\int^{v_+}_{v_-}(\dot{\theta_+}r^3+3\theta_+r^2\dot{r})\dot{r'}dv \nonumber \\
&=\frac{1}{2}\int^{v_+}_{v_-}\frac{\theta_+^2-R_{vv}}{r^2}(R_{\theta\theta}-1
-r^2\theta_+\theta_-  )z_1^4dv,  
\end{align}
where we used Eq.~(\ref{condition:f}) in the last line. 
Substituting Eq.~(\ref{formula_2}) into Eq.~(\ref{formula_1}) and using Eq.~(\ref{condition:f}) , one obtains 
\begin{align}
\label{formula_3}
& \int^{v_+}_{v_-}z_1^4\,\p_v(\p_u(g_{(2)vv})(0,v))dv \nonumber \\
&=\int^{v_+}_{v_-}z_1^4\theta_+\left(\theta_+\dot{f'}-\frac{\theta_-}{2}R_{vv}\right)dv \nonumber \\
&+\frac{1}{2}\int^{v_+}_{v_-}\frac{\theta_+^2-R_{vv}}{r^2}(R_{\theta\theta}-1
-r^2\theta_+\theta_- )z_1^4dv \nonumber \\
&=-\int^{v_+}_{v_-}z_1^4
\left[\theta_+^2R_{uv}+\frac{1}{2r^2}R_{vv}(R_{\theta\theta}-1) \right]dv. 
\end{align}
Similarly, the integration of $h_{(4)vv}$ in Eq.~(\ref{Delta_u_4}) is rewritten in terms of expansions and the Ricci curvature 
of the boundary spacetime~(\ref{boundary:d=4}) as 
\begin{align}
& \int^{v_+}_{v_-} z_1^4\ln z_1\,h_{(4)vv}(0,v)dv \nonumber \\
&=-\frac{1}{12}\int^{v_+}_{v_-}z_1^4\frac{\theta_+}{r}\ddot{r'}dv+\frac{1}{6}\int^{v_+}_{v_-}z_1^4\frac{\dot{r}r'\ddot{r}}{r^3}dv
+\frac{1}{12}\int^{v_+}_{v_-}z_1^4\left(\frac{\dot{r}^2}{r^2}+\frac{\ddot{r}}{r} \right)\dot{f'}dv \nonumber \\
&=\frac{1}{96}\int^{v_+}_{v_-}z_1^4\left[\theta_+^2\left(\frac{R_{\theta\theta}}{r^2}-2R_{uv} \right)
-\theta_+^2\left(\theta_+\theta_-+\frac{1}{r^2} \right)-2\theta_-\theta_+ R_{vv}+4R_{vv}R_{uv}    \right]dv
\end{align}
by integration by parts.



\begin{thebibliography}{99}
\bibitem{Tipler1978}
F.~J.~Tipler, 
``Energy conditons and spacetime singularities'', 
Phys.\ Rev.\ D {\bf 17}, no. 10, 2521 (1978). 

\bibitem{WaldYurtsever91}
R.~M.~Wald adn U. Yurtsever, 
Phys.\ Rev.\ D {\bf 44}, 403 (1991). 

\bibitem{GrahamOlum07}
N.~Graham and K.~D.~Olum, 
Phys.\ Rev.\ D {\bf 76}, 064001 2007).


\bibitem{FlanaganWald96}.  
E.~E.~Flanagan and R.~M.~Wald, 
Phys.\ Rev.\ D {\bf 54}, 6233 (1996). 



\bibitem{Faulkner:2016mzt} 
  T.~Faulkner, R.~G.~Leigh, O.~Parrikar and H.~Wang,
  ``Modular Hamiltonians for Deformed Half-Spaces and the Averaged Null Energy Condition,''
  JHEP {\bf 1609}, 038 (2016)
  doi:10.1007/JHEP09(2016)038
  [arXiv:1605.08072 [hep-th]].


\bibitem{Hartman:2016lgu} 
  T.~Hartman, S.~Kundu and A.~Tajdini,
  ``Averaged Null Energy Condition from Causality,''
  JHEP {\bf 1707}, 066 (2017)
  doi:10.1007/JHEP07(2017)066
  [arXiv:1610.05308 [hep-th]].

\bibitem{Kelly:2014mra} 
  W.~R.~Kelly and A.~C.~Wall,
  ``Holographic proof of the averaged null energy condition,''
  Phys.\ Rev.\ D {\bf 90}, no. 10, 106003 (2014)
  Erratum: [Phys.\ Rev.\ D {\bf 91}, no. 6, 069902 (2015)]
  doi:10.1103/PhysRevD.90.106003, 10.1103/PhysRevD.91.069902
  [arXiv:1408.3566 [gr-qc]].


\bibitem{Maldacena:1997re}
  J.~M.~Maldacena,
  ``The Large N limit of superconformal field theories and supergravity,''
  Int.\ J.\ Theor.\ Phys.\  {\bf 38}, 1113 (1999)
  [Adv.\ Theor.\ Math.\ Phys.\  {\bf 2}, 231 (1998)]
  doi:10.1023/A:1026654312961, 10.4310/ATMP.1998.v2.n2.a1
  [hep-th/9711200].
  



\bibitem{Rosso2019}
F.~Rosso,  
``Global aspects of conformal symmetry and the ANEC in dS and AdS'', 
JHEP {\bf 2020}, 186 (2020) 186  [arXiv:1912.08897 [hep-th]]. 

\bibitem{Rosso2020}
F.~Rosso,  
``Achronal averaged null energy condition for extremal horizons and (A)dS'', 
JHEP 07 (2020) 023
[arXiv:2005.0647 [hep-th]]. 

\bibitem{Urban:2009yt} 
  D.~Urban and K.~D.~Olum,
  ``Averaged null energy condition violation in a conformally flat spacetime,''
  Phys.\ Rev.\ D {\bf 81}, 024039 (2010)
  doi:10.1103/PhysRevD.81.024039
  [arXiv:0910.5925 [gr-qc]].
 
 
\bibitem{IMM:2019} 
  A.~Ishibashi, K.~Maeda and E.~Mefford,
  ``Achronal averaged null energy condition, weak cosmic censorship, and AdS/CFT duality,''
  Phys.\ Rev.\ D {\bf 100}, no. 6, 066008 (2019)
  doi:10.1103/PhysRevD.100.066008
  [arXiv:1903.11806 [hep-th]].

\bibitem{Gao:2000ga} 
  S.~Gao and R.~M.~Wald,
  ``Theorems on gravitational time delay and related issues,''
  Class.\ Quant.\ Grav.\  {\bf 17}, 4999 (2000)
  doi:10.1088/0264-9381/17/24/305
  [gr-qc/0007021].

\bibitem{IIM:2020}
N.~Iizuka, A.~Ishibashi, and K.~Maeda, 
``Conformally invariant averaged null energy condition from AdS/CFT,'' 
JHEP {\bf 03} (2020) 161 [arXiv:  1911.02654]


\bibitem{deHaro:2000vlm} 
  S.~de Haro, S.~N.~Solodukhin and K.~Skenderis,
  ``Holographic reconstruction of space-time and renormalization in the AdS / CFT correspondence,''
  Commun.\ Math.\ Phys.\  {\bf 217}, 595 (2001)
  doi:10.1007/s002200100381
  [hep-th/0002230].
  
\bibitem{Hickling:2015tza}
   A.~Hickling and T.~Wiseman,
  ``Vacuum energy is non-positive for (2 + 1)-dimensional holographic CFTs"
   Class. Quant. Grav. {\bf 33}, 045009 (2016)
   doi:10.1088/0264-9381/33/4/045009     
   [arXiv:1508.04460 [hep-th]].
 
 \bibitem{Fischetti:2016vfq}
   S.~Fischetti, A.~Hickling and T.~Wiseman,
 ``Bounds on the local energy density of holographic CFTs from bulk geometry"
   Class. Quant. Grav. {\bf 33}, 225003 (2016)
   doi:10.1088/0264-9381/33/22/225003
   [arXiv:1605.00007 [hep-th]]. 

\bibitem{BrownHenneaux}
J.~D. Brown and M.~Henneaux, 
``Central Charges in the Canonical Realization of Asymptotic Symmetries: an
Example from Three-Dimensional Gravity'' 
Comm. Math. Phys. 104 (1986) 207.

\bibitem{Borde1987}
A.~Borde, 
``Geodesic Focusing, energy conditions and singularities,'' 
Class. Quant. Grav. {\bf 4}, 343 (1987). 


\bibitem{BTZ}
M.~Banados, C.~Teitelboim, J.~Zanelli, ``{The Black hole in three-dimensional space-time}",
Phys.\ Rev.\ Lett. {\bf 69}, 1849 (1992)
[hep-th/9204099]

\bibitem{Hayward:1994bu} 
S.A.~Hayward, ``{Gravitational energy in spherical symmetry}",
Phys. Rev. D {\bf 53}, 1938--1949, (1996)
 doi:10.1103/PhysRevD.53.1938 
 [gr-cq/9408002] 

\bibitem{WaldIshibashi04}
A.~Ishibashi and R.~M.~Wald, 
``Dynamics in Non-Globally-Hyperbolic Static Spacetimes III: Anti-de Sitter Spacetime,'' 
Class. Quant. Grav. {\bf 21}, 2981 (2004). 


\end{thebibliography}
\end{document}